\documentclass[aps,prl,twocolumn,groupedaddress]{revtex4}
\usepackage{graphicx}
\begin{document}

\title{Radiation Driven Inflation}

\author{Aharon Davidson}
\email[Email: ]{davidson@bgu.ac.il}
\author{Ilya Gurwich}
\email[Email: ]{gurwichphys@gmail.com}

\affiliation{Physics Department, Ben-Gurion University,
Beer-Sheva 84105, Israel}                 

\date{April 15, 2008}

\begin{abstract}
	A novel scalar field free approach to cosmic inflation is presented.
	The inflationary Universe and the radiation dominated Universe
	are shown, within the framework of unified brane cosmology,
	to be two different phases governed by one and the same energy density.
	The phase transition of second order (the Hubble constant exhibits
	a finite jump) appears naturally and serves as the exit mechanism.
	No re-heating is needed.
	The required number of e-folds is achieved without fine tuning.
\end{abstract}	

\maketitle   

The remarkable idea that our Universe has undergone, in its
very earliest stages of evolution, a phase of exponential
expansion\cite{inflation1,inflation2} is widely accepted as the
solution to the horizon, flatness, and magnetic  monopole puzzles.
The inflationary Universe scenario\cite{review}, which has gained
strong experimental support from the detailed observations of the
Cosmic Microwave Background radiation\cite{WCMB}, is now
considered part of the standard hot Big Bang cosmology.
With this in mind, it is highly frustrating that the physical mechanism
underlying inflation is essentially obscure.
The conventional ad-hoc theoretical prescription of the inflationary
scenario invokes a scalar field of some sort, in analogy to the Higgs
field introduced in particle physics.
The accompanying potential is carefully engineered to address certain
desired features of inflation.
It is not clear what degrees of freedom are collectively represented
by this so-called inflaton, and what actually determines the shape of
its model dependent potential.
The beginning and the end the inflationary era are theoretical challenges
by themselves, with the major goal being the production of a sufficient
number $N_{e}\simeq 60$ of $e$-folds.
While the beginning may resemble a spontaneous creation, by means
of quantum nucleation\cite{nucleation,DKL1}, the end traditionally occurs
once the inflaton starts oscillating around the absolute minimum of
the scalar potential.

In this paper, within the framework of unified brane gravity\cite{DG},
we present a novel approach to cosmic inflation.
The inflationary Universe and the subsequent radiation
dominated Universe are shown to be two different phases governed
by one and the same brane energy density.
It differs from other models of inflation in that it does
not involve scalar fields and/or scalar potentials\cite{ScalarFree}.
Alternatively, the model forces the brane energy/momentum tensor,
predominantly in the inflationary phase, to consist of radiation
and surface tension, which are essential standard cosmological
ingredients.
Furthermore, the model offers a natural built-in exit mechanism,
implemented by means of a second order phase transition.
The Universe exiting the inflationary era is then necessarily radiation
dominated and hot. 
In turn, no re-heating\cite{inflation2,reheat} is needed, neither as
a graceful end of Guth's inflation, nor as the arena for recreating
matter and fill the Universe with radiation.
In addition, the required number of $e$-folds emerges naturally
without fine tuning.

Dirac\cite{Dirac}, in his 'Extensible Model of the Electron', has paved
the way for a consistent brane variation.
He was concerned with the fact that the 'linearity of the variation' may
be in jeopardy.
Rephrasing Dirac, 'a tiny deformation of the brane corresponding
to the brane being pushed a little to the right will not be minus the
variation corresponding to the brane being pushed a little (equally)
to the left, on account of the left and right bulk sections not being
a smooth continuation of each other'.
To bypass the problem, a general curvilinear coordinate system has
been invoked, such that in the new so-called Dirac frame, \emph{the
location of the brane does not change during the variation process}.
Imposing Dirac's prescription on modern brane theories based on an
action principle, such as the Collins-Holdom\cite{CH} model, which
unifies the familiar Randall-Sundrum\cite{RS} and the
Dvali-Gabadadze-Porrati\cite{DGP} models, and assuming a discrete
$Z_{2}$ symmetry on simplicity grounds, the corresponding brane
field equations (see Ref.\cite{DG} for the derivation) take the form
\begin{equation}
	\begin{array}{c}
	 \displaystyle{\frac{1}{4\pi G_5}
	\left(K_{\mu\nu}-g_{\mu\nu}K\right)  =}  \vspace{4pt}\\
	  \displaystyle{= \frac{1}{8\pi G_4}
	\left(R_{\mu\nu}-\frac{1}{2}
	g_{\mu\nu}R\right)+
	T_{\mu\nu}+\lambda_{\mu\nu} ~, }
	\end{array}
	\label{FieldEq}
\end{equation}
where $G_{5(4)}$ denote the gravitational coupling constants in the
bulk (brane), respectively.
In addition to the conventional terms (the Israel\cite{IJC} junction
term, the Einstein tensor associated with the scalar curvature $R$
on the brane, and the physical energy-momentum tensor of the
brane $T_{\mu\nu}=\delta {\cal L}_{matter}/\delta g^{\mu\nu}$),
unified brane gravity gives furthermore rise to $\lambda_{\mu\nu}$.
The latter tensor consists of Lagrange multipliers associated with the
fundamental induced metric constraint
$g_{\mu\nu}(x)-g_{MN}(y(x))y^{M}_{,\mu}y^{N}_{,\nu}=0$.
It has been proven that $\lambda_{\mu\nu}$ is conserved, and that
its contraction with the extrinsic curvature vanishes
\begin{equation}
	\lambda^{\mu\nu}_{~;\nu}=0~, \quad
	\lambda_{\mu\nu}K^{\mu\nu}=0 ~.
	\label{ubgbasic}
\end{equation}
As is evident from the above field equations, $\lambda_{\mu\nu}$
serves as a geometrical (embedding originated) contribution to the
total energy-momentum tensor of the brane, and as such, may have
far reaching gravitational consequences.
It is thus crucial to make sure that, although deviating from the RS
approach, the GR limit is still there.
Cosmological analysis has been shown\cite{DG} to reproduce the
GR limit.
On top of it, by analyzing the weak field perturbations caused
by the presence of matter on the brane, we have already
recovered\cite{GDr} the 4-dimensional Newton force law.

Within a cosmological framework, eq.(\ref{FieldEq}) can be
integrated out\cite{DG}, giving rise to a novel constant of
integration $\omega$, the finger print of unified brane cosmology.
The corresponding FRW equation which governs the evolution of
the $4$-dimensional brane can be conveniently written
in the form
\begin{equation}
	\frac{8\pi G_{N}}{3}\rho_{eff}(a)\equiv
	\frac{\dot{a}^{2}+k}{a^{2}}=
	\frac{\Lambda_{5}}{6}+\xi^2 (a) ~.
	\label{FRW}
\end{equation}
If the only energy/momentum source in the bulk is a negative
cosmological constant $\Lambda_{5}<0$, then $\xi(a)$ has been
shown\cite{DG} to be a solution of the cubic equation
\begin{equation}
	P(\xi) \equiv \frac{3\xi^{2}}{8\pi G_{4}}+
	\frac{3\xi}{4\pi G_{5}}+
	\frac{\Lambda_{5}}{16\pi G_{4}}-\rho(a)+
	\frac{~\omega}{\sqrt{3}\xi a^{4}} =0~.
	\label{cubic}
\end{equation}
Eqs.(\ref{FRW},\ref{cubic}) thus generalize the familiar RS,
DGP and CH cosmologies, which are recovered at the
$\omega\rightarrow 0$ limit.
The Regge-Teitelboim\cite{RT} Cordero-Vilenkin\cite{stealth}
stealth Universe models are manifest at the $G_{5}\rightarrow \infty$
limit.
Like in all brane cosmologies\cite{branecos,braneinflation}, it is
crucial to notice that $\rho_{eff}(a)$ defined in eq.(\ref{FRW}) is
no longer the physical energy density on the brane, but rather a
function of it.
It is $\rho(a)=T^0_0$ which serves as the physical energy density
on the brane.

\begin{figure}[ht]
	\includegraphics[scale=1]{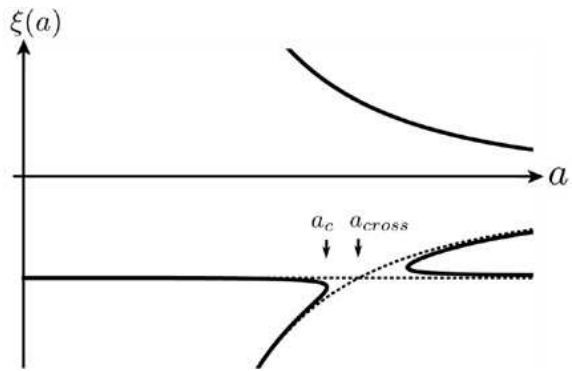}
	\caption{\label{xi}
	Given $\rho(a)=\sigma +A/a^{4}$,
	two roots of cubic eq.(\ref{cubic}) exhibit a hyperbolic structure.
	The intersection point of the two asymptotes (the straight
	horizontal one associated with eternal inflation) sets a
	natural scale for terminating inflation.}
\end{figure}

Let us first focus attention on the special case of \emph{eternal}
inflation, which clearly requires eq.(\ref{cubic}) to admit an
\emph{exact} $a$-independent solution.
Such a solution, conveniently written as $\xi (a)=\omega/\sqrt{3}A$,
with $A$ being a constant, as depicted by the straight horizontal
(dashed) asymptote in fig.(\ref{xi}), corresponding to a (positive
by assumption) cosmological constant
\begin{equation}
	\Lambda_{0}=\frac{1}{2}\Lambda_{5}
	+\frac{\omega ^{2}}{A^{2}} ~,
\end{equation}
may exist for some conventional energy density $\rho(a)$.
The serendipitous observation now being  that $\rho(a)$ must
solely consist (as hinted first in ref.\cite{DKL2}) of radiation
accompanied by a particular amount of surface tension
\begin{equation}
	\begin{array}{l}
	\displaystyle{\rho(a)=\frac{A}{a^{4}}+\sigma_{0}} ~,
	\vspace{8pt}\\
	\displaystyle{\sigma_{0}=\frac{1}{8\pi}
	\left(\frac{\omega^{2}}{G_{4}A^{2}}+
	\frac{2\sqrt{3}\omega}{G_{5}A}+
	\frac{\Lambda_{5}}{2 G_{4}}
	\right)} ~.
	\end{array}
	\label{rho}
\end{equation}
The 'no-ghost' condition $A>0$ is then naturally adopted.

Counter intuitively, however, even 'eternal' inflation cannot last
forever in our model.
The slightest deviation of the physical surface tension $\sigma$
from the particular $\sigma_{0}$ value, will expose, as shown in
fig.(\ref{xi}), the hyperbolic structure of the $\xi(a)$ roots.
The amount of finite inflation, however, is insensitive to the value
of $\sigma$.
The intersection point of the two asymptotes (the straight one
is associated with eternal inflation), as determined directly from
eq.(\ref{cubic}), with eq.(\ref{rho}) imposed, sets a natural FRW scale
$a_{cross}$ for terminating inflation, namely
\begin{equation}
	a_{cross}^{4}=
	\frac{4\pi G_4 A^2}{\sqrt{3}\omega \left(
	\displaystyle{\frac{G_4}{G_5}+\frac{\omega}{\sqrt{3}A}}
	\right)} ~.
	\label{acritical}
\end{equation}
The early Universe cosmological constant $\Lambda_{0}$
is also not sensitive to the value of the surface tension $\sigma$.
In fact, on realistic grounds (to be revealed soon), $\sigma_{0}$ will be
traded for the Randall-Sundrum surface tension
\begin{equation}
	\sigma_{RS}=\frac{3}{4\pi G_{5}}
	\sqrt{-\frac{\Lambda_{5}}{6}} ~,
	\label{simgmaRS}
\end{equation}
with the price being a finite, yet sufficient, amount of radiation
driven inflation.

At very short scale factors, the situation may appear to be confusing.
While the energy density $\rho(a)$ clearly explodes as $a\rightarrow 0$,
the effective energy density $\rho_{eff}(a)$ stays finite in this limit.
What this means is that, in the absence of an ultra-violet cutoff (included
in particular is the $k=0$ flat case), the inflationary era can in fact start
at an arbitrarily small scale.
In turn, the total number of $e$-folds can become arbitrarily large.
In such a case, $a_{enter}$ will mark the edge of validity (set by the Planck
scale) of our classical field equations.
If $k>0$, on the other hand, the cosmological scale factor marking the
entrance of inflation gets fixed by $a_{enter}=\sqrt{3k/\Lambda_{0}}$.
Interpreted as spontaneous creation of a closed baby Universe,
such an entrance is presumably governed by Hawking-Hartle\cite{nucleation}
no-boundary proposal, or alternatively, by
Davidson-Karasik-Lederer\cite{DKL1} brane nucleation.

A great number of physical processes may occur during the later
stages of cosmic evolution.
One of which, for example, reflecting the fact that the massive
particles have eventually cooled down, is the appearance of a dust
term $B/a^{3}$ in the energy/momentum tensor.
Our interest is focused, however,  on the early Universe exiting the
inflationary era, and not on the very late Universe.
In this respect, it remains to be seen whether radiation driven inflation,
based on $\rho(a)=\sigma+A/a^{4}$
is capable of spontaneously inducing an exit mechanism, and whether
the emerging Universe happens to correspond to the physical one we
know of.
If we insist on $\sigma>0$, to assure a positive Newton's constant
at the general relativistic limit (as well as brane stability), the cubic
eq.(\ref{cubic}) admits three real roots such that
$\xi_{-}(\infty)<\xi_{0}(\infty)=0<\xi_{+}(\infty)$ as $a\rightarrow \infty$.
Only the largest of these roots, namely
\begin{equation}
	\xi_{+}(\infty)= -\frac{G_{4}}{G_{5}}
	+\sqrt{\frac{8\pi G_{4}}{3}\sigma
	+\frac{G_{4}^{2}}{G_{5}^{2}}
	-\frac{\Lambda_{5}}{6}} ~,
\end{equation}
is capable of supporting an adjustably tiny cosmological constant
$\Lambda_{\infty}\simeq 0$, as required by the present Universe (the
other two roots lead to a negative and a positive definite $\Lambda_{\infty}$,
respectively).
This naturally calls for the familiar Randall-Sundrum fine-tuning
of the surface tension, given by eq.(\ref{simgmaRS}), with the subsequent
Collins-Holdom identification\cite{CH} of Newton's constant
\begin{equation}
	\frac{1}{G_{N}}=\frac{1}{G_{4}}+
	\frac{1}{G_{5}}\sqrt{-\frac{6}{\Lambda_{5}}} ~.
\end{equation}
To probe the nature of the Universe at the post inflationary era, we
expand $\xi_{+}(a)$ to learn that it is radiation dominated
\begin{equation}
	\frac{\dot{a}^{2}+k}{a^{2}} \simeq \frac{8\pi G_{N}}{3}
	\left(1-\frac{\omega}{A}\sqrt{-\frac{2}{\Lambda_{5}}}\right)
	\frac{A}{a^{4}} ~,
	\label{radiation}
\end{equation}
provided the enhancement factor (in parentheses) is positive.

At this stage, an apparent contradiction is encountered.
On the one hand, inflation has been shown to single out the $\xi_0(a)$
branch at the small-$a$ regime, whereas it is the $\xi_{+}(a)$
branch which is required for the large-$a$ regime.
One must thus closely follow the time evolution of the three roots,
with the FRW scale factor serving as the evolution parameter.
Notice that the role of a finite $a$ is to add the linear piece
$(-A\xi+\frac{~1}{\sqrt{3}} \omega)/a^{4}$ to the asymptotic
($a\rightarrow \infty$) expression of $P(\xi)$.
There are four cases to examine, corresponding to the different
regions on the $\xi$-axis where $\omega/\sqrt{3}A$ can be located.
A careful analysis reveals that
(i) There is no $\omega$ for which $\xi(a)$ would analytically
evolve from $\xi_{0}(a)$ at small-$a$ to $\xi_{+}(a)$ at large-$a$,
(ii) There exists a range of parameters, namely
\begin{equation}
	\frac{~\omega}{\sqrt{3}A} <\xi_{-}(\infty)=
	-\frac{2G_{4}}{G_{5}}-
	\sqrt{-\frac{\Lambda_{5}}{6}}~,
	\label{range}
\end{equation}
for which $\xi_{0}(a)$ at small-$a$ \emph{must} connect
with $\xi_{+}(a)$ at large-$a$. The connection is established
by means of a second order phase transition, and
(iii) Within the above range, the radiation enhancement
factor in eq.(\ref{radiation}) is positive, in fact $>2$. The exiting
Universe is thus \emph{necessarily} radiation dominated. 
\begin{figure}[ht]
	\includegraphics[scale=1]{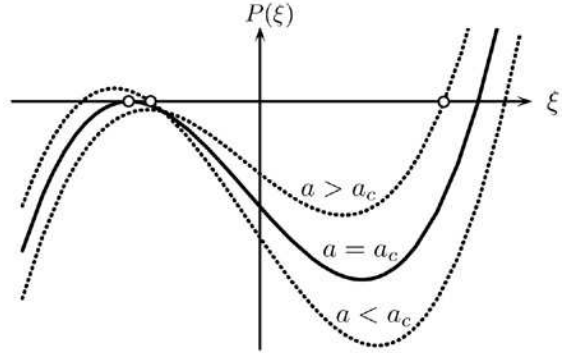}
	\caption{\label{polynomial}
	The roots $\xi(a)$ of the cubic equation $P(\xi)=0$ serve
	to express the effective energy density $\rho_{eff}(a)$ as
	a function of the physical energy density $\rho(a)$.
	Responding to the finite jump $\xi_{0}(a_c -\epsilon)$
	$\rightarrow$ $\xi_{+}(a_c +\epsilon)$, the Universe undergoes
	an inflation $\rightarrow$ radiation domination phase transition.}
\end{figure}

\begin{figure}[ht]
	\includegraphics[scale=1]{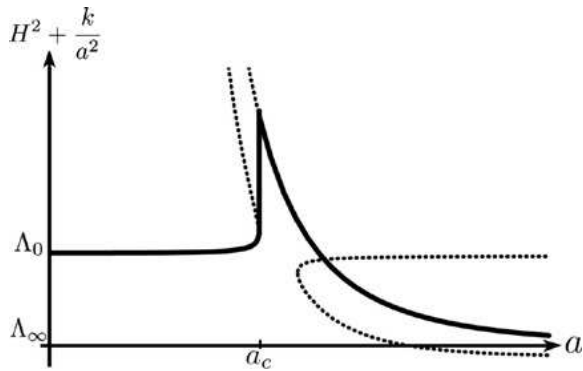}
	\caption{\label{rhoeff} 
	The early Universe phase transition, connecting the
	inflationary ($a<a_c$) and the radiation dominated
	($a>a_c$) eras, is characterized by a finite jump in the
	Hubble constant $H$.
	Remarkably, the two phases share the one and the same
	physical energy density $\rho(a)=\sigma_{RS}+A/a^4$.}
\end{figure}

Plotted in fig.(\ref{polynomial}), with $a$ serving as the parameter,
is the master cubic polynomial $P(\xi)$.
As long as the FRW scale factor is sub-critical, that is $a<a_{c}$,
there exist three real solutions, the central of which $\xi_{0}(a)$
is recognized as the inflation oriented solution.
Once $a$ approaches criticality, $\xi_{0}(a)$ and $\xi_{-}(a)$
merge, and are about to mutually disappear (becoming complex)
as $a$ crosses the $a_{c}$ barrier.
The only left over real solution is then $\xi_{+}(a)$.
Responding to the finite jump
$\xi_{0}(a_{c})=\xi_{-}(a_{c})\rightarrow \xi_{+}(a_{c})$,
the Universe undergoes a second order phase transition.
While the FRW scale factor $a(t)$ remains continuous, the
(positive) Hubble constant exhibits a sudden \emph{finite
increase} (note that $\ddot{a}\rightarrow +\infty$ when
nearing $a_c$ from below).
As in any phase transition in physics, however, it is expected
that fluctuations will smoothen the jump ($\ddot{a}<+\infty$).
The effective energy/momentum tensor is characterized by
a typical critical behavior.
Expanding near (below) the critical point, one finds 
\begin{equation}
	\begin{array}{l}
	 \rho_{eff}(a) \simeq \alpha
	 -\beta(a_{c}-a)^{1/2}  \vspace{4pt} \\ 
	 P_{eff}(a)  \simeq
	 \displaystyle{-\frac{\beta a_{c}}{6}(a_{c}-a)^{-1/2}} 
	\end{array} 
\end{equation}
for some positive constants $\alpha,\beta$.
$P_{eff}(a)$ is the corresponding effective pressure.
The effective energy density $\rho_{eff}(a)$, characterized by
a finite jump at the critical scale, is plotted in fig.(\ref{rhoeff}).
The shape of the graph (not necessarily its physics) highly
reminds us of the specific heat as a function of $1/T$ (which in
fact is the scale factor) in $\lambda$-transition of liquid helium.
At any rate, as $a$ keeps growing in the radiation dominated
phase, one will face once again three real roots, but this time
for a change, real $\xi_{+}(a)$ is never lost.
It is interesting to remark that, due to the Hysteresis-like nature
of the evolution, time reversibility is not respected.
In other words, a shrinking radiation dominated Universe will
\emph{not} undergo deflation.

To estimate the total number $N_{e}$ of $e$-folds generated by
radiation driven inflation, one needs to calculate the critical value
$a_{c}$ of the FRW scale factor.
On simplicity grounds (nothing to do with fine tuning), let us do
it in the limit where $\omega$ approaches the right edge of the
physically allowed range specified by eq.(\ref{range}), that is for
$\omega \rightarrow \sqrt{3}A \xi_{-}(\infty)$, which is fully
equivalent to setting $\sigma_{RS} \rightarrow\sigma_{0}$.
This simple limit is characterized by the fact that the two branches
$\xi_{-}(a)$ and $\xi_{0}(a)$ intersect, with the intersection point
serving as the origin of a local hyperbolic structure generated once
$\omega$ deviates from the limit value.
For $\omega \le \sqrt{3}A \xi_{-}(\infty)$, we have
$a_{c}\le a_{cross}$.

We now claim that, up to the $\Lambda_{\infty}\simeq 0$ fixing,
which is a severe fine-tuning problem by itself, our model is free
of the conventional inflation-oriented fine-tuning.
To address the naturalness issue of the various ratios in the theory,
we define the dimensionless positive constant
\begin{equation}
	\gamma = \frac{G_{5}}{G_{4}}
	\sqrt{-\frac{\Lambda_{5}}{6}}>0~.
\end{equation}
Combining the entrance and exit scales, we can estimate the
total number of $e$-folds, and find
\begin{equation}
	N_{e}\le\frac{1}{4}\ln\frac{16\pi (\gamma+1)G_{N}A 
	\Lambda_{0}}{9\gamma(\gamma+2) k^{2}} ~.
	\label{Ne}
\end{equation}
It is customary to associate the entrance with the Planck scale
(the alternative being the GUT scale), that is
$\Lambda_{0}\simeq 1/t_{Planck}^{2}$ , and further require,
taking into account today's relative suppression of radiation
versus matter densities,
$A \simeq 10^{-4} \rho_{c}t_{Hubble}^{4}$, where 
$\rho_{c}=3/8\pi G_{N}t_{Hubble}^{2}$ denotes today's
critical energy density.
This brings us to the region around
\begin{equation}
	N_{e} \le
	\frac{1}{2}\ln\frac{10^{-2}t_{Hubble}}{t_{Planck,GUT}}
	= \left\{
	\begin{array}{ll}
	69  &  ~Planck \vspace{4pt}\\
	65  &  ~GUT     
	\end{array}
	\right.
	 ~.
\end{equation}
Although the result is not that sensitive to the value of
$\gamma$ (for example, changing $\gamma$ by a factor
of hundred will only contribute $\pm 1$), it is fully consistent
with $\gamma$ being roughly ${\cal O}(1)$, thus emphasizing
the naturalness of radiation driven inflation

To summarize, the general idea of radiation driven inflation
may seem to be self contradictory at the first glance.
After all, it takes a deviation from general relativity to allow
the physical energy density $\rho(a)$, stemming from the
brane matter Lagrangian, to differ from $\rho_{eff}(a)$,
the effective source of the FRW geometry.
Within the framework of unified brane gravity, the interplay
of these two energy densities is taken one step beyond the
Randall-Sundram model, when noticing that
$\rho_{eff}(a)\simeq const$ actually dictates
$\rho(a)\sim 1/a^4$ at small scales (note that at such small
scales all particles, massive as well, would be ultra relativistic
and thus radiation like).
This opens the door for the fascinating possibility that the
inflationary Universe and the subsequent radiation dominated
Universe are in fact two different phases governed by one
and the same physical energy/momentum tensor.
Furthermore, unlike other models of inflation, brane inflation
models included, the present one does not invoke ad hoc
scalar fields.
As a consequence,
the Universe must have been hot before as well as after the
phase transition.
No re-heating is thus in order.
Furthermore, notice that unlike in early models of
inflation, $a_{exit}$ and hence the number of $e$-folds does
not depend here on initial conditions.
Obviously, since a scalar potential is not a part of our theory,
conditions such as 'starting almost at rest at the top of the hill'
and 'slow-roll' are irrelevant.
Altogether, up to the usual $\Lambda_{\infty}\simeq 0$,
radiation driven inflation is fine-tuning free.
Needless to say, a number of theoretical questions are still to be
addressed in our inflation model, most notably the issue of
the phase transition.
In particular, it is crucial to understand the behavior of the
matter fields under such a transition.
The research of fluctuations is also in order.
As in any phase transition in physics, one expects the
fluctuations to create "islands" of the second phase, that tend
to expand and eventually consume the entire universe, thereby
"smoothening" the phase transition.

\section*{Acknowledgment}
We would like to cordially thank Professors Kameshwar Wali, Ray 
Volkas and Eduardo Guendelman for enlightening discussions,
and Shimon Rubin for constructive comments.


\begin{thebibliography}{}
\bibitem{inflation1}
	A.H. Guth, Phys. Rev. D23, (1981) 347.
\bibitem{inflation2}
	A.D. Linde, Phys. Lett. B108, (1982) 389.
	A. Albrecht and P.J. Steinhardt, Phys. Rev. Lett. 48, (1982) 1220.
	A.D. Linde, Phys. Lett. B129, (1983) 177.
\bibitem{review}
	A.D. Linde, Rept. Prog. Phys. 47, (1984) 925.
	K.A. Olive, Phys. Rept. 190, (1990) 307.
	J.E. Lidsey, A.R. Liddle, E.W. Kolb, E.J. Copeland, T. Barreiro,
		and M. Abney, Rev. Mod. Phys. 69, (1997) 373.
	D.H. Lyth and A. Riotto, Phys. Rep. 314, (1999) 1.
\bibitem{WCMB}
	D.N. Spergel, et al., ApJS, 170, (2007) 377.
	M. Tegmark, et al. (SDSS collaboration), Phys. Rev. D74,
	(2006) 123507.
	L. Alabidi and D.H. Lyth, J. Cosmol. Astropart. Phys. 0608,
	(2006) 013.
\bibitem{nucleation}
	S.W. Hawking and I.G. Moss, Phys. Lett. 110B, (1982) 35.
	J.B. Hartle and S.W. Hawking, Phys. Rev. D28, (1983) 2960.
 	A.D. Linde, Sov. Phys. JETP 60, (1984) 211.
    	A. Vilenkin, Phys. Lett. 117B, (1982) 25; Phys. Rev. D30, (1984) 509.
\bibitem{DKL1}
	A. Davidson, D. Karasik and Y. Lederer, Class. Quant. Grav.
	16, (1999) 1349.	
\bibitem{DG}	
	A. Davidson and I. Gurwich, Phys. Rev. D74, (2006) 044023.
\bibitem{ScalarFree}
	L.H. Ford, Phys. Rev. D40, 967 (1989);
	S. Watson, M.J.Perry. G.L. Kane and F.C. Adams, JCAP 0711,
	017 (2007).
\bibitem{reheat}
	L. Kofman, A.D. Linde, and A.A. Starobinsky, Phys. Rev. Lett.
	73, (1994) 3195.
	Y. Shtanov, J. Trachen and R. Brandenberger, Phys. Rev. D51,
	(1995) 5438.
\bibitem{Dirac}
	P.A.M. Dirac, Proc. Roy. Soc. London \textbf{A268}, (1962) 57.
\bibitem{CH}	
	H. Collins and B. Holdom, Phys. Rev. D62, (2000) 105009.
	H. Collins and B. Holdom, Phys. Rev. D62, (2000) 124008.	
\bibitem{RS}
	L. Randall and R. Sundrum, Phys. Rev. Lett. 83, (1999) 3370;
	Phys. Rev. Lett. 83, (1999) 4690.
\bibitem{DGP}
	G. Dvali, G. Gabadadze, and M. Porrati, Phys. Lett. B484, (2000) 112;
	Phys. Lett. B485, (2000) 208.
\bibitem{IJC}
	W. Israel, Nuovo Cimento \textbf{B44}, (1966) 1.
\bibitem{GDr}
	    I. Gurwich and A. Davidson "Linearized Unified Brane Gravity"
	    (in preparation).
\bibitem{RT}
	T. Regge and C. Teitelboim, in Proc. Marcel Grossman (Trieste),
	(1975) 77.
	S. Deser, F.A.E. Pirani, and D.C. Robinson, Phys. Rev. D14,
	(1976) 3301.
	A. Davidson and D. Karasik, Mod. Phys. Lett. A13, (1998) 2187.
	A. Davidson, Class. Quan. Grav. 16, (1999) 653.
\bibitem{stealth}
	R. Cordero and A. Vilenkin, Phys. Rev. D65, (2002) 083519.
\bibitem{branecos}
	E. Papantonopoulos, Lect. Notes Phys. 592, (2002) 458.
	P. Brax and C. van de Bruck, Class. Quant. Grav. 20, (2003) R201.
	D. Langlois, Prog. Theor. Phys. Supp. 148, (2003) 181.
\bibitem{braneinflation}
	G.R. Dvali and S.H.H. Tye, Phys. Lett. B450, (1999) 72.
	R. Maartens, D. Wands, B.A. Bassett and I.Heard,
		Phys. Rev. D62, (2000) 041301.		
\bibitem{DKL2}
	A. Davidson, D. Karasik, and Y. Lederer, Phys. Rev. D72,
	(2005) 064011.

\end{thebibliography}
\end{document}